# THE HABITABILITY OF GJ 357D: POSSIBLE CLIMATES AND OBSERVABILITY

L. Kaltenegger[1,2], J. Madden[1,2], Z. Lin[1,2], S. Rugheimer[1,3], A. Segura[4], R. Luque[5,6], E. Pallé[5,6], N. Espinoza[7]

[1] Carl Sagan Institute, Cornell University, Space Science Institute, 14850 Ithaca, NY, USA, lkaltenegger@astro.cornell.edu, Tel: +1-607-255-3507
[2] Cornell University, Space Science Institute, 14850 Ithaca, NY, USA,
[3] Oxford University, Oxford, OX1 3DW, UK
[4] Instituto de Ciencias Nucleares, Universidad Nacional Autónoma de México, México
[5] Instituto de Astrofisica de Canarias, 38205 La Laguna, Tenerife, Spain
[6] Departamento de Astrofisica, Universidad de La Laguna, 38206, La Laguna, Tenerife, Spain
[7] Max Planck Institute, Koenigstuhl 17, 69117 Heidelberg, Germany

ABSTRACT

The GJ 357 system harbors 3 planets orbiting a bright, nearby M2.5V star at 9.44pc. The innermost planet GJ 357b (TOI-562.01) is a hot transiting Earth-size planet with Earth-like density, which receives about 12 times the irradiation Earth receives from the Sun, and was detected using data from TESS. Radial velocities discovered two more planets in the system at 9.12 (GJ 357c) and 55.6 days (GJ 357d), with minimum masses of 3.59±0.50 and 6.1±1 Earth masses, and an irradiation of 4.4 and 0.38 Earth's irradiation, respectively. GJ 357d receives slightly less stellar irradiation than Mars does in our own Solar System, which puts it in the Habitable Zone for its host star. GJ 357d could not have been detected with TESS and whether it transits remains an open question.

Here we model under what conditions GJ 357d could sustain surface habitability and present planetary models as well as synthetic transmission, reflection and emission spectra for a range of models for GJ 357d from water worlds to Earth-like models. With Earth-analog outgassing rates, GJ 357d would be a frozen rocky world, however with an increased $CO_2$ level, as would be expected if a geological cycles regulates $CO_2$ concentration like on Earth, the planet models show temperate surface conditions. If we can detect a transit of GJ 357d, it would become the closest transiting, potentially habitable planet in the solar neighborhood. Even if GJ 357d does not transit, the brightness of its star makes this planet in the Habitable Zone of a close-by M star a prime target for observations with Extremely Large telescopes as well as future space missions.

1. INTRODUCTION

To date more than 4.000 exoplanets have been discovered, providing a first glimpse of the diversity of other worlds (e.g. reviews by Winn & Fabrycky 2015, Udry & Santos 2007). Several of these planets receive irradiation from their host star that is similar to Earth, which could provide liquid water and habitable surface environments for rocky planets or moons (see e.g. review Kaltenegger 2017, Batalha 2014, Kane et al 2016).

For several hundred exoplanets both mass and radius are known and thus we can estimate the mean density of the planet, which can be used to derive its composition and compare it to planets in our own Solar System. Figure 1 shows the diversity of small known exoplanets with the error bars on the measurements. We chose 4 $R_{Earth}$ and 20 $M_{Earth}$ as limits in Figure 1 to include the most massive known rocky planets, so called *Super-Earths* like Kepler-10c, a planet with about 18 $M_{Earth}$ and 2.3 $R_{Earth}$ (Dumusque et al. 2014) consistent with a rocky composition (see Zeng & Sasselov 2013). However several known exoplanets with masses down to 1 $M_{Earth}$, have radii corresponding to gas planets or so called *Mini-Neptunes* e.g. Kepler11-f has a mass between 1.1 and 5 $M_{Earth}$, but a radius of 2.6 $R_{Earth}$ (Lissauer et al. 2011).

Colored lines in Figure 1 show exoplanetary density models for different composition from Iron (100% Fe) to Earth-like ($MgSiO_3$ (rock)) to a pure $H_2O$ composition (100% $H_2O$), encompassing the densest to lightest rocky composition for an exoplanet (following Zeng, Sasselov & Jacobsen 2016). Earth and Venus are shown in black for reference. Figure 1 shows that gas planets can have masses down to 1 $M_{Earth}$, while planets with masses up to 18 $M_{Earth}$ can also be rocky, making the mass of a planet a very weak constraint on its composition. The color-coding in Fig. 1 indicates the effective surface temperature of the host star, which provides additional insight into the composition of small mass planets by host star spectral type.

GJ 357 b is the only planet in this system where a transit was detected so far, its radius is 1.217 ± 0.084 Earth radii, $R_{Earth}$, and its mass 1.84 ± 0.31 $M_{Earth}$. GJ 357 b shows a mean density like Earth, while both GJ 357c and GJ 357d currently only have minimum masses from RV measurements, thus we don't know their radii or bulk composition. Figure 1 shows that in the region of minimum mass range for these two planets a wide variety of exoplanets from Super-Earths to Mini-Neptuns have been detected. The effective surface temperature of the host star indicates most planets in these mass ranges for M dwarfs, shown in red in Fig.1, are consistent with a rocky composition.

While a wide range of compositions and atmosphere is possible for GJ 357 d, we focus on the interesting case that this planet could have a rocky composition (see Fig.1). We model a range of scenarios for GJ 375 d, from a planet model similar to Earth, which leads to a rocky radius of 1.75 (Zeng et al 2013) and gravity of twice of Earth's to the limiting case for the largest rocky planet of this mass, a water world, which leads to a rocky radius of 2.4 $R_{Earth}$ and a gravity of 1.16 times Earth's for GJ 375 d.

We can use the incident stellar flux planets receive to compare planetary environments: Present-day Venus e.g. receives 1.9 times the Solar Flux at Earth's orbit, $S_0$, present-day Mars receives 0.4 $S_{Earth}$. Any rocky planet that receives more flux than present-day Venus is empirically too hot to be habitable. GJ 357 b and GJ 357 c receive about 13 times and 4.4 times the Earth's irradiation ($S_{Earth}$), respectively. For comparison Venus receives about 1.9 $S_{Earth}$ and Mercury about 6.5 $S_{Earth}$. Thus, both planets should have undergone a runaway greenhouse stage as proposed for Venus' evolution and lost their water. On the other hand, GJ 357 d receives an irradiation of 0.38 $S_{Earth}$, which places it inside the Habitable Zone (HZ), in a location comparable to Mars in the Solar System, making it a very interesting target for further atmospheric observations.

The HZ is a concept that is used to guide remote observation strategies to characterize potentially habitable worlds: It is defined as the region around one or multiple stars in which liquid water could be stable on a rocky planet's surface (e.g., Kasting et al. 1993, Kaltenegger & Haghighipour 2013, Kane

& Hinkel 2013, Kopparapu et al. 2013, Ramirez & Kaltenegger 2017), facilitating the remote detection of possible atmospheric biosignatures. The width and orbital distance of a given HZ depends to a first approximation on two main parameters: incident stellar flux and planetary atmospheric composition. The incident stellar flux depends on stellar luminosity, stellar spectral energy distribution, the planet's orbital distance (semimajor axis) and eccentricity of the planetary orbit. The warming due to atmospheric composition depends on the planet's atmospheric makeup, energy distribution, and resulting albedo and greenhouse warming.

A star's radiation shifts to longer wavelengths with cooler surface temperatures, which makes the light of a cooler star more efficient at heating an Earth-like planet with a mostly $N_2$-$H_2O$-$CO_2$ atmosphere (see e.g. Kasting et al. 1993). This is partly due to the effectiveness of Rayleigh scattering, which decreases at longer wavelengths. A second effect is the increase in NIR absorption by $H_2O$ and $CO_2$ as the star's spectral peak shifts to these wavelengths, meaning that the same integrated stellar flux that hits the top of a planet's atmosphere from a cool red star warms a planet more efficiently than the same integrated flux from a star with a higher effective surface temperature (See Fig.2). Stellar luminosity as well as the SED change with stellar spectral type and age, which influences the orbital distance at which an Earth-like planet can maintain climate conditions which allow for liquid water on its surface (see review by Kaltenegger 2017).

Fig.2 shows the *empirical HZ, which* is based on observations in our own Solar System (see Kasting et al. 1993). The inner edge of this empirical HZ, the so-called *Recent Venus limit*, is based on the observation that Venus may have had liquid water on its surface until about 1 billion years ago, consistent with atmospheric D/H ratio measurements suggesting a high initial water endowment (Donahue and Pollack,1983). Note that the inner limit is not well known because of the lack of a reliable geological surface history of Venus beyond about 1 billion years due to resurfacing of the stagnant lid, which allows for the possibility of a liquid surface ocean, however it does not stipulate a liquid ocean surface. At that time the Sun was ~ 8% less bright than today, yielding a solar flux equivalent equal to 1.76 present-day Solar irradiance at Earth's orbit ($S_{Earth}$). The empirical outer edge for the HZ, the so-called *Early Mars limit*, is based on observations suggesting that Mars did not have liquid water on its surface at or before 3.8 billion years ago. At that time the solar flux was about 25% lower or equal to about 0.32 $S_0$. The corresponding orbital distances in our Solar System are 0.75 AU (Recent Venus limit) and 1.77 AU (Early Mars limit), respectively for present solar luminosity, excludes present-day Venus and includes present-day Mars. Note that being in the HZ does not necessarily mean that a planet is habitable, and in-depth follow-up spectral observations of their atmospheres are needed to characterize planets and search for signs of life (see review Kaltenegger 2017)

Fig. 2 shows the known transiting M star planetary systems with planets with less than 3 $R_{Earth}$ in terms of orbital semimajor axis of their planets as well as the contours of the Habitable Zone. It compares the GJ 357 planetary system with another well-known planet system with 7 Earth-size planets, TRAPPIST-1, which is at a similar distance from the Sun as GJ 357 and another interesting target for observations because it harbors 3 planets in the HZ (Guillon et al 2017) and an additional 4th planet added in the Volcanic Habitable Zone (Ramirez & Kaltenegger 2017). GJ 357 d (0.38 ± 0.01 $S_{Earth}$) receives comparable stellar irradiation to TRAPPIST-1 f (0.35 ± 0.02 $S_{Earth}$) and also orbits in the outer part of the HZ.

Most exoplanets with small minimum masses orbit in the HZ of dim M dwarfs (e.g. Udry et al. 2007; Anglada-Escudé et al. 2013, 2016; Tuomi & Anglada-Escudé 2013; Dittmann et al. 2017; Gillon et al. 2017). Thus the brightness of GJ 357 makes this system a very interesting target for observations and atmospheric characterization of the planets' atmospheres. If future observations e.g. with CHEOPs (Broeg *et al.*, 2013) can detect a transit of GJ 357d, it would become the closest transiting planet in the HZ, allowing for in-depth studies of its atmosphere. However even if GJ 357 d does not transit, the brightness of its star makes this planet in the Habitable Zone of a close-by M star a prime target for ground and space based observations. For transiting terrestrial planets around the closest stars, the James Web Space Telescope scheduled for launch in 2021 (e.g. Gardner et al., 2006, Clampin *et al.*, 2009; Deming et al., 2009, Kaltenegger & Traub, 2009, Barstow & Irwin 2016), as well as upcoming ground-based telescopes (e.g. Snellen et al.2013, Roedler & Morales 2014), might be able to detect biosignatures in a rocky planet's atmosphere for planets around the closest stars. The ELTs, will focus on observations in the visible, but also have capabilities in the NIR to IR like the METIS instrument at the ELT. Observations can also characterize planetary atmospheres if the planet does not transit their host star due to the known orbital movement and resulting radial velocity shift (see e.g. Birksby et al. 2019). Several space mission concepts to characterize Earth-like planets are currently being designed e.g. by NASA's science and technology definition teams, but no concept has been selected yet. Different concepts like stars-shades and coronagraphs are designed to take spectra of extrasolar planets with the ultimate goal of remotely detecting atmospheric signatures to characterize nearby Super-Earths and Earth-like planets, enable comparative planetology beyond our Solar System and search for signs of life on other worlds.

Signs of life that modify the atmosphere or the surface of a planet and thus can be remotely detectable are key to finding life on exoplanets or exomoons (see e.g. review Kaltenegger 2017). Observations of our Earth with its diverse biota function as a Rosetta Stone to identify habitats. Some atmospheric species exhibiting noticeable spectral features in our planet's spectrum as a result directly or indirectly from biological activity: the main ones are $O_2$, $O_3$, $CH_4$, $N_2O$ and $CH_3Cl$. Any biosignature need to be analyzed critically for potential geological sources under conditions different from those on Earth (see e.g. Kasting et al 2013, Kaltenegger 2017, Walker et al. 2018; Meadows et al 2018). For prebiotic chemistry additional chemicals have been proposed as atmospheric signatures to look for (see e.g. Ranjan et al. 2017, Rimmer et al 2018). Spectroscopy can reveal the presence of specific molecules and atoms by identifying their characteristic energy transitions. Section 2 discusses our models, section 3 shows our results and section 4 discusses and concludes our paper.

2. METHODS

In this paper we focus on GJ 357 d, a planet with a minimum mass of 6.1 +/- 1 Earth masses, $M_{Earth}$ (see Luque et al 2019).

This translates into a rocky planetary radius of 1.75 $R_{Earth}$ assuming rocky composition and 2.4 $R_{Earth}$ assuming pure ice composition, which is the limiting case for the largest core radius for a rocky planet (see Zeng & Sasselov 2013 for details). This translates into a surface gravity of about twice Earth's surface gravity for a 1.75 Earth radius model and 1.16 Earth's gravity for a 2.4 Earth radius model.

We use the spectra model for the M2V active stars model (described in Rugheimer et al. 2015) as the host star input spectra, which has a similar effective surface temperature to GJ 357. The UV stellar spectrum is based on IUE data (see Rugheimer et al. 2015). We model four different types of atmospheres here for GJ 357 d for rocky composition, a radius of 1.75 $R_{Earth}$ and a gravity of 2 times Earth's gravity as well as for a water world composition, a radius of 2.4 $R_{Earth}$ and a gravity of 1.16 times Earth's gravity:

Two atmospheric models assume Earth-analog outgassing rates for surface pressure of 1 bar and 2 bar, two more models, one anoxic and one oxic atmosphere assumes increased greenhouse effect from $CO_2$ $CH_4$, and $H_2$ concentrations added until the planet's average surface temperature is at least 273 for a 2 bar and 5 bar surface pressure, respectively.

For the two scenarios with Earth-analog outgassing ratios (see Rugheimer et al. 2014) but different surface pressure of 1 bar and 2 bars, we keep the planetary outgassing rates for $H_2$, $CH_4$, CO, $N_2O$, and $CH_3Cl$, and mixing ratios of $O_2$ at 0.21 and $CO_2$ at $3.55 \times 10^{-6}$, with a varying $N_2$ concentration that is used as a fill gas to reach the set surface pressure of the model following Segura et al. (2005). Note that by keeping the outgassing rates constant, higher surface pressure atmosphere models initially have slightly lower mixing ratios of chemicals with constant outgassing ratios than lower surface pressure models.

We model an additional anoxic case for each planet core model, where we increase $CO_2$ concentration to maintain an average surface pressure above freezing for the planets – assuming the planet also has a similar geological cycle like Earth's carbonate silicate cycle, which stabilizes the surface temperature and regulates CO2 concentration in the atmosphere over geological timescales (see e.g. review Kaltenegger 2017). This could provide liquid water and habitable surface conditions on the planet. For the anoxic atmospheres we assume a mixing ratio of $CO_2$ at 0.1, $CH_4$ at 0.048, and $H_2$ at 0.16. For the 1.75 Earth radii case this yields a surface temperature of 273K and for the 2.4 Earth radii case a surface temperature of 288K. We include $H_2$ as an additional greenhouse gas due to the larger masses of these planets and it's possible role in the heating of early Mars at a similar effective insolation (see e.g. Ramirez et al., 2014).

The model for a warm oxic atmosphere for GJ 357d employs an atmospheric mixing ratio of 0.1 $O_2$ and a biotic methane flux of $8.57 \times 10^{10}$ molecules/cm$^2$/s consistent with a 0.1 $O_2$ atmosphere during the Proterozoic (Olson et al. 2016). A mixing ratio of 0.8 $CO_2$ and a surface pressure of 5 bars results in a surface temperature of 288K. Note that different atmospheric compositions (e.g. different mixing ratios of $CO_2$, $CH_4$ and $H_2$) can maintain temperatures for liquid water on the surface of GJ 357d and our model only shows one possibility for reference.

For this study, we use a 1D climate and photochemistry code, Exo-Prime, a coupled 1D radiative-convective atmosphere code developed for rocky exoplanets coupled to a line-by-line Radiative transfer code, which generates the spectra in different viewing geometries (see e.g. Kaltenegger et al 2009, Kaltenegger & Sasselov 2010, Rugheimer et al 2013, 2015, Rugheimer & Kaltenegger 2018) to model the Earth-like atmospheres and a 1D atmospheric model developed for the study of habitability of GJ581 d (Kaltenegger et al. 2011).

EXO-Prime simulates both the effects of stellar radiation on a planetary environment and the planet's outgoing spectrum. We model an altitude range in the atmosphere that extends upwards to a minimum of 70km with 50 height layers. We use a geometrical model in which the average 1D global atmospheric model profile is generated using a plane-parallel atmosphere, treating the planet as a Lambertian sphere, and setting the stellar zenith angle to 60 degrees to represent the average incoming stellar flux on the dayside of the planet (see also Schindler & Kasting 2000). The temperature in each layer is calculated from the difference between the incoming and outgoing flux and the heat capacity of the atmosphere in each layer. If the lapse rate of a given layer is larger than the adiabatic lapse rate, it is adjusted to the adiabatic rate until the atmosphere reaches equilibrium.

We use a two-stream approximation (see Toon et al. 1989), which includes multiple scattering by atmospheric gases, in the visible/near IR to calculate the shortwave fluxes. Four-term, correlated-$k$ coefficients parameterize the absorption by $O_3$, $H_2O$, $O_2$, and $CH_4$ (Kasting and Ackerman, 1986). A fixed relative humidity is assumed following Manabe and Wetherald (1967). The tropospheric lapse rate follows a moist adiabat (Kasting, 1988) that takes into account $CO_2$ and $H_2O$ condensation. For all the models $N_2$ concentration is calculated to fill out the atmosphere after the concentrations of the other chemical species have been set up. In the thermal IR region, a rapid radiative transfer model (RRTM) calculates the longwave fluxes for oxic atmospheres and IR absorption model for high $CO_2$ atmospheres (Pavlov et al 2000, Haqq-Misra et al. 2008).

Clouds are not explicitly calculated. The photochemistry code, originally developed by Kasting et al. (1985) solves for 55 chemical species linked by 220 reactions using a reverse-Euler method (see Segura et al. 2010, and references therein). The anoxic atmosphere model is explained in Segura et al (2007), Haqq-Misra et al. 2008 and Kaltenegger et al. (2011).

The photochemical model is stationary, and convergence is achieved when the following criteria are fulfilled: the production and loss rates of chemical species are balanced which results in a steady state for the chemical concentrations, and the initial boundary conditions, such a surface mixing ratios or surface fluxes, are met. Photolysis rates for various gas-phase species are calculated using a δ two-stream routine (Toon et al., 1989) that accounts for multiple scattering by atmospheric gases and by sulfate and hydrocarbon aerosols. One important feature of the high $CO_2$ model is its ability to keep track of the atmospheric hydrogen budget, or redox budget. H and $H_2$ escape was simulated by assuming an upward flux at the diffusion limited rate (Walker, 1977).

The radiative transfer model used to compute planetary spectra is based on a model originally developed for trace gas retrieval in Earth's atmospheric spectra (Traub & Stier 1976) and further developed for exoplanet transmission and emergent spectra (Kaltenegger et al. 2007; Kaltenegger & Traub 2009; Kaltenegger 2010; Kaltenegger & Sasselov 2010; Kaltenegger et al. 2013). In this paper, we model Earth's transmission, reflected and thermal emission spectra using 21 of the most spectroscopically significant molecules ($H_2O$, $O_3$, $O_2$, $CH_4$, $CO_2$, OH, $CH_3Cl$, $NO_2$, $N_2O$, $HNO_3$, CO, $H_2S$, $SO_2$, $H_2O_2$, NO, ClO, HOCl, $HO_2$, $H_2CO$, $N_2O_5$, and HCl).

For the reflected and emitted spectra, we use a Lambert sphere as an approximation for the disk integrated planet in our model. The surface of our model planet corresponds to Earth's current surface of 70% ocean, 2% coast, and 28% land. The land surface consists of 30% grass, 30% trees, 9% granite, 9% basalt, 15% snow, and 7% sand. Surface reflectivities are taken from the USGS Digital Spectral Library and the ASTER Spectral Library (following Kaltenegger et al. 2007). For our larger water worlds we assume a similar overall surface albedo for ease of comparison between our models. Note that the change in surface albedo between a liquid water surface – low albedo and stronger absorption of incoming irradiation – or a pristine frozen world – high reflectivity, would in addition influence the planet's climate. However it is an ongoing scientific discussion on how pristine an icy surface would remain on such a world. Without further input, we chose to maintain a similar overall surface albedo of about 0.16 for all our models.

For the transmission spectrum, we trace the light from the star through individual layers in the atmosphere and then combined the spectra to the overall transmission spectra of the planet as discussed in detail in Kaltenegger & Traub 2009 and Betremieux & Kaltenegger 2014.

We calculate the spectrum at high spectral resolution with several points per line width. The line shapes and widths are computed using Doppler and pressure broadening on a line-by-line basis for each layer in the model atmosphere. The overall high-resolution spectrum is calculated with 0.1 $cm^{-1}$ wavenumber steps. The figures are shown smoothed to a resolving power of 700 using a triangular smoothing kernel. The spectra may be binned further for comparison with proposed future spectroscopy missions designs to characterize Earth-like planets. We previously validated EXO-Prime from the VIS to the IR using data from ground and space (see e.g. Kaltenegger et al. 2007; Kaltenegger & Traub 2009; Rugheimer et al. 2013).

All of oxygenic simulations used a fixed mixing ratio of 355ppm for $CO_2$ and 21% $O_2$. For the sample anoxic model we set $CO_2$ = 1.0 x $10^{-1}$ and $H_2$ to 1.9 x $10^{-4}$. The $N_2$ mixing ratio is set to be a fill gas such that the total surface pressure is 1 or 2 bar and a fixed upper boundary of $10^{-7}$ bar (a minimum of 70 km).

RESULTS

GJ 357 d has a minimum mass of 6.1 +/- 1 $M_{Earth}$ (see Luque et al 2019). Assuming the minimum mass is the real planet's mass, this translates into a rocky planetary radius of 1.75 $R_{Earth}$, assuming rocky composition and 2.4 $R_{Earth}$ assuming pure ice composition, which is the limiting case for the largest core radius for a rocky planet (see Zeng & Sasselov 2013 for details). This translates into a surface gravity of about twice Earth's surface gravity for a 1.75 Earth radius model and 1.16 Earth's gravity for a 2.4 $R_{Earth}$ model.

GJ 357 d receives 0.38 times Earth's irradiation.

We model three different types of atmospheres here for GJ 357 d for i) a rocky composition, a radius of 1.75 $R_{Earth}$ and a gravity of 2 times Earth's gravity as well as ii) a water world composition, a radius of 2.4 $R_{Earth}$ and a gravity of 1.16 times Earth's gravity. These two atmospheric models assume Earth-analog outgassing rates for surface pressure of 1 bar and 2 bar. We also model an anoxic atmosphere as a third example, where we increase $CO_2$ concentration to a mixing ratio of $10^{-1}$, under which the planet's average surface temperature is above freezing.

GJ 357 d received similar flux to Mars in our own Solar System. However it is more massive than Mars and if we assume geological activity, similar to Earth, an increase in atmospheric $CO_2$ at lower stellar irradiation is expected. On a geologically active planet like Earth the geochemical carbonate-silicate cycle stabilizes the long-term climate and atmospheric $CO_2$ content, depending on the surface temperature: $CO_2$ is continuously outgassed and forms carbonates in the presence of surface water, which then get subducted and $CO_2$ gets outgassed again. Farther from the star, the lower stellar irradiance would create a cooler surface temperature on a planet, thus linking the orbital distance to atmospheric $CO_2$ concentrations levels: $CO_2$ should be a trace gas close to the inner edge of the HZ but a major compound in the outer part of the HZ with several bar of $CO_2$ (e.g. Walker 1981). Because the outer limit of the HZ is based on the assumption that atmospheric $CO_2$ will buildup and warm the planet, an adequate $CO_2$ mixing ratio in the atmosphere is needed to maintain surface temperatures above freezing for planets on the outer part of the HZ. We increase $CO_2$ levels in our anoxic model runs, to a mixing ratio to $10^{-1}$, which provides surface conditions above freezing for both planet models, a rocky core composition as well as an icy core composition.

The average surface temperature for the sample models are 211 K (1 bar, 1.75 $R_{Earth}$), 215 K (2 bar, 1.75 $R_{Earth}$), 215 K (1 bar, 2.4 $R_{Earth}$), 221 K (2 bar, 2.4 $R_{Earth}$) for the models assuming Earth-like outgassing rates without and increase in $CO_2$. For our anoxic atmosphere model with an increase in $CO_2$ and $H_2$ levels the average surface temperature is 273 K (2 bar, 1.75 $R_{Earth}$, anoxic) and 288 K (2 bar, 2.4 $R_{Earth}$, anoxic). For our oxic high $CO_2$ atmosphere the surface temperature is 278 K (5 bar, 1.75 $R_{Earth}$, oxic).

Figure 3 shows the temperature as well as water, ozone, methane, and $H_2$ mixing ratios versus height in the model atmospheres for GJ 357 d. Note that some oxygen and ozone also builds up in the atmosphere of terrestrial planets around M stars without a biological source as expected (see e.g. Domagal-Goldman, et al. 2014, Wordsworth, R., & Pierrehumbert, R. 2014, Hu, R. et al 2012).

MODEL SPECTRA FOR GJ 357D

Encoded in the planet's transmission, reflection and emission spectra is the information of the chemical make-up of a planet's atmosphere and if the atmosphere is transparent, the emergent spectrum also carries some information about surface properties for emergent flux. That makes light from a planet a crucial tool to characterize it. The presence or absence of spectral features will indicate similarities or differences of the atmospheres of terrestrial exoplanets from Earth, and their astrobiological potential.

We show synthetic spectra for upcoming space and ground telescopes in Fig. 4 and Fig.5 with a spectral resolution

($\lambda/\Delta\lambda$) of 700 as a sample resolution, like the high resolution setting for the NIRSPEC instrument on JWST. While different spectral resolution is envisioned for different instruments, 700 gives a good sample overview of how the spectral features will appear. We keep the spectral resolution constant for the spectra shown in the figures to allow easy cross comparison and indicate soe commonly used filters in Astronomy.

Note that all spectra are available online (carlsaganinstitute.org/data) in high resolution (run at 0.1 cm$^{-1}$ wavenumber, which corresponds to a spectral resolution ($\lambda/\Delta\lambda$) of 100,000 at 1μm). The high-resolution spectra can then be smeared to any required resolution.

Several teams have discussed how to observe spectral features of temperate rocky planets (e.g. Kaltenegger et al 2009. Palle et al 2009, Snellen et al. 2012, Morley et al. 2017) and a full model for the different instruments available on JWST as well as the ELTs should be run, once it is determined whether the planet transits, which would constrain the radius and thus the gravity and models for the planet.

Reflection and Emission spectra for GJ 357d

Figure 4 shows the reflected flux (top) and (bottom) emission spectra for our models for GJ 357 d as sample spectra for JWST as well as ELTs for direct imaging or secondary eclipse measurements. We include the relative reflection spectra (Fig. 4 middle) here to show the effect of the incident starlight on the detectable chemical species. As for any spectral features, the amount of chemicals needed to show a feature varies, depending on the geometry of observations as well as atmospheric composition and temperature.

The depth of spectral features in reflected light is dependent on the abundance of a chemical as well as the incoming stellar radiation at that wavelength. The sample spectra for GJ 357 d shows strong atmospheric features in reflected light (Fig.4 top) from 0.7 to 4μm, especially for the warmer anoxic sample atmospheres for $H_2O$ at e.g. 1.4μm and 1.9 μm, and a smaller $CH_4$ features at 1.7μm, and 2.4 μm, Earth-like atmospheres show $O_2$ at 0.76 μm. The region between 2 and 5 microns is not shown in Fig. 4 because the flux from the planet in reflection and emission is extremely low.

In thermal emission, the depth of spectral features depends on the abundance of a chemical as well as the temperature difference between the emitting/absorbing layer and the continuum. In the IR, the strongest atmospheric features in our models for GJ 357 d from 4 to 20μm are $O_3$ at 9.6μm, $CO_2$ at 15 μm, $H_2O$ at 6.4μm and 17μm and $CH_4$ at 7.7μm. The IR shows the difference in surface temperature of the planet models strongly with both anoxic models being run with $CO_2$ concentrations that warm the planet above freezing compared to the other cooler surface, thus showing a larger emission than the planet models with colder surfaces, therefor to see the features of the cooler planets we did not include the hotter anoxic models in Fig 4 but they are available online. As shown in Fig. 4 high amounts of $CO_2$ dominate the infrared emission spectrum, as previously shown in model spectra for a warm habitable Gl 581d (Kaltenegger et al. 2011). The low overall flux also shows that the layer that can be observed is not the surface, which would be a hotter black body temperature, and thus, that the atmosphere can not be probed to the ground for the high $CO_2$, 5 bar atmospheric model. Thus the ozone feature in the emission spectra is not detectable in this model, however the oxygen feature in the reflection spectra can still be observed (see also Kaltenegger et al. 2011).

Note that for both reflected light and emitted light we did not show the influence of the size of the planet on the overall flux to be able to compare the spectra in one figure. However the overall flux of the planet model, which assumes ice composition for the core is about twice as large as the model for the rocky core, due to the difference in corresponding radius (1.75 $R_{Earth}$ versus 2.4 $R_{Earth}$).

For an Earth-like biosphere, the main detectable atmospheric chemical signatures that in combination could indicate habitability are $O_2/O_3$ in combination with $CH_4$ or $N_2O$, and $CH_3Cl$. Reflection as well as the emission spectra of GJ 357 d show both features for Earth-like atmospheres (see e.g. reviews by DesMarais et al 2007, Kasting et al. 2014, Kaltenegger 2017, Schwieterman et al. 2018).

TRANSMISSION SPECTRA FOR GJ 357D

Even though no transit for GJ 357d has been detected yet, we also model transmission spectra for out model cases. Gravity, mean molecular weight and temperature of a planet affects its transmission spectrum through the pressure profile of its atmosphere, and hence its atmospheric absorption profile. For an ideal gas atmosphere in hydrostatic equilibrium, the pressure, p, varies with the altitude, z, as $d \ln(p) = -1/H \, dz$, where H is the atmospheric scale height defined as $H = kT/\mu g$, where $k$ is Boltzmann's constant and $T, \mu$ and $g$ are the local (i.e., altitude dependent) temperature, mean molecular mass, and gravity, respectively (Figure 5 top row). The transit signal, the flux difference, $\Delta F$, depends on the planet's radius divided by the star's radius squared (Figure 5 middle & bottom row).

Figure 5 (top panels) shows the comparison of the effective height of the atmosphere in transit for our model planets in the top panel: increasing gravity of the planet (twice Earth's gravity for a radius of 1.75 $R_{Earth}$ versus 1.16 time Earth's gravity for a 2.4 $R_{Earth}$) translates into a reduced transmission height with increased gravity. Increasing temperature increases the scale height and transmission signal, note that the anoxic cases for GJ 357 d were run to produce a surface temperature above freezing, thus generate a warmer surface as well as atmospheric temperature due to increased greenhouse warming than the Earth-like models, where we did not increase greenhouse gases above present-day Earth. Note that these spectra are not showing the effect of the planet's size on the overall planet's flux for consistency. Planets with radius 2.4 $R_{Earth}$ versus 1.75 $R_{Earth}$ have a larger area and thus will appear brighter.

Figure 5 shows that the largest signal is generated by the lower gravity model of the 2.4 $R_{Earth}$ planet, with the largest signal being generated by the warmest, denser 2 bar surface pressure atmosphere. All 2.4 $R_{Earth}$ atmospheres at similar surface pressure produce larger signals than the 1.75 $R_{Earth}$ models, because of the lower gravity assumed for the 2.4 $R_{Earth}$ models.

The sample transit spectra for GJ 357 d shows strong atmospheric features in reflected light (Fig.5 top) from 0.7 to 5μm, especially for the warmer anoxic and oxic sample atmospheres for $H_2O$ at e.g. 1.4μm and 1.9 μm, and a smaller $CH_4$ features at 1.7μm, and 2.4 μm. Oxic atmospheres show $O_2$ at 0.76 μm. In the IR, the strongest atmospheric features in our models for GJ 357 d from 5 to 20μm are $CO_2$ at 15 μm, $H_2O$ at 6.4μm and 17μm and $CH_4$ at 7.7μm. For the Earth-like atmospheres containing oxygen, the $O_3$ at 9.6μm is visible. Note that due to the small size of its host star compared to the planet, assuming a rocky composition, the flux difference is favorable for future observations.

Given the brightness of the host star, these targets could be prime targets for JWST observations. Using PandExo (Batalha et al 2017) to estimate the achievable precision per transit observation, we find that as few as 4 transits could allow for the detection of key atmospheric species such as $CO_2$. Note that this target is currently beyond the brightness limits of the NIRISS instrument[1] and would benefit from the implementation of a high-efficiency read mode as outlined in Batalha et al. (2018).

5. CONCLUSIONS & DISCUSSION

We present planetary models as well as synthetic transmission, reflection and emission spectra for a range of atmospheres for the newly discovered planet GJ 357 d, which orbits in the Habitable Zone of its host star. The host star harbors two more known planets, the innermost is a confirmed transiting planet with a mean density similar to Earth. While it is still unknown if GJ 357d transits, the brightness of its host star makes this planet in the Habitable Zone of a close-by M star a prime target for ground- and space based atmosphere characterization. Assuming the minimum mass of GJ 357 d is the real planet's mass, this translates into a rocky planetary radius of 1.75 $R_{Earth}$, assuming rocky composition and 2.4 $R_{Earth}$ assuming pure ice composition, which is the limiting case for the largest core radius for a rocky planet. We model under what conditions GJ 357 d could sustain liquid water and surface habitability for a range of different atmospheric conditions from Earth-like to anoxic atmospheres, for rocky to water worlds.

GJ 357 d receives 0.38 times Earth's irradiation, similar to Mars in our own Solar System. We model three different types of atmospheres here for GJ 357 d for i) a rocky composition and ii) a water world composition. For Earth-analog outgassing rates for different surface pressures from 1 to 2 bar, the surface temperature remains below freezing. However geological active worlds, like our Earth, are expected to build up $CO_2$ concentrations due to the feedback of the carbonate-silicate cycle. We model oxic and anoxic atmospheres as three examples, where we increase $CO_2$ concentration, so that the planet's average surface temperature is above freezing.

The sample reflection, emission and transmission spectra show features of a wide range of chemicals, $H_2O$, $CO_2$, $CH_4$, and $O_3$ and $O_2$ for Earth-like atmospheres from the Visible to Infrared wavelength (0.4 to 20 μm), which would indicate habitability for observations with upcoming telescopes.

**Acknowledgements:** Special Thanks to Nikole Lewis and Morgan Saidel.

---

[1] https://jwst-docs.stsci.edu/near-infrared-imager-and-slitless-spectrograph/niriss-predicted-performance/niriss-bright-limits)

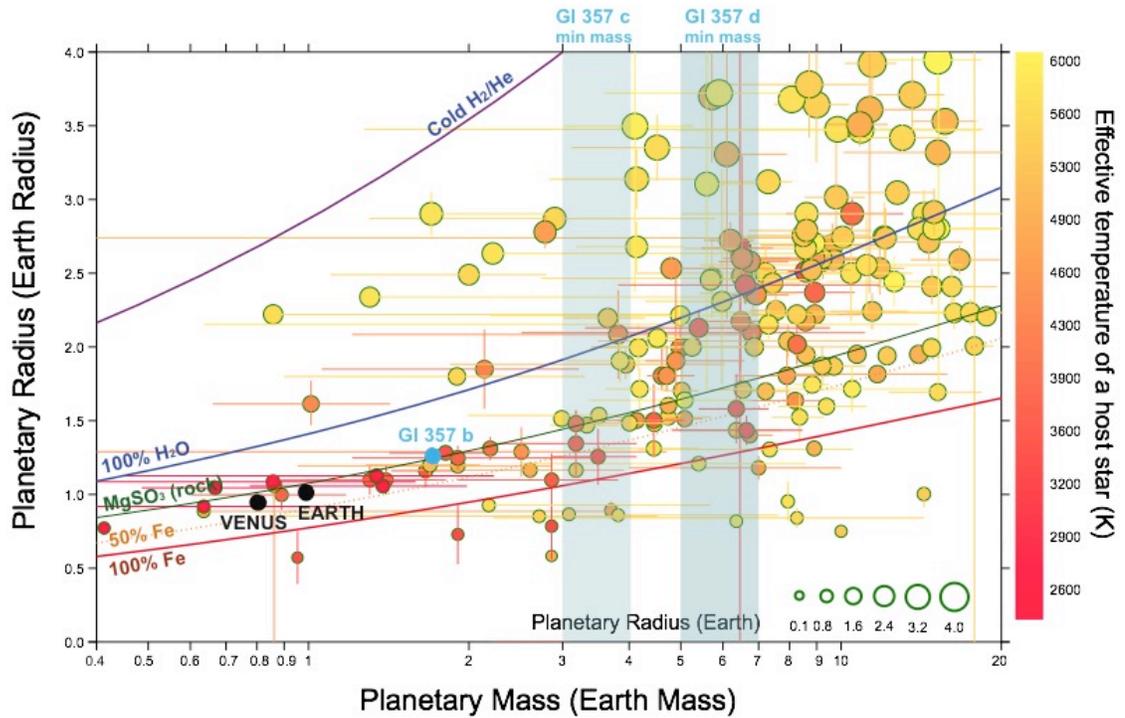

**Figure 1**: Mass radius diagram of detected exoplanets which have both mass and radius measurements (data exoplanet.eu, June 19 2019)

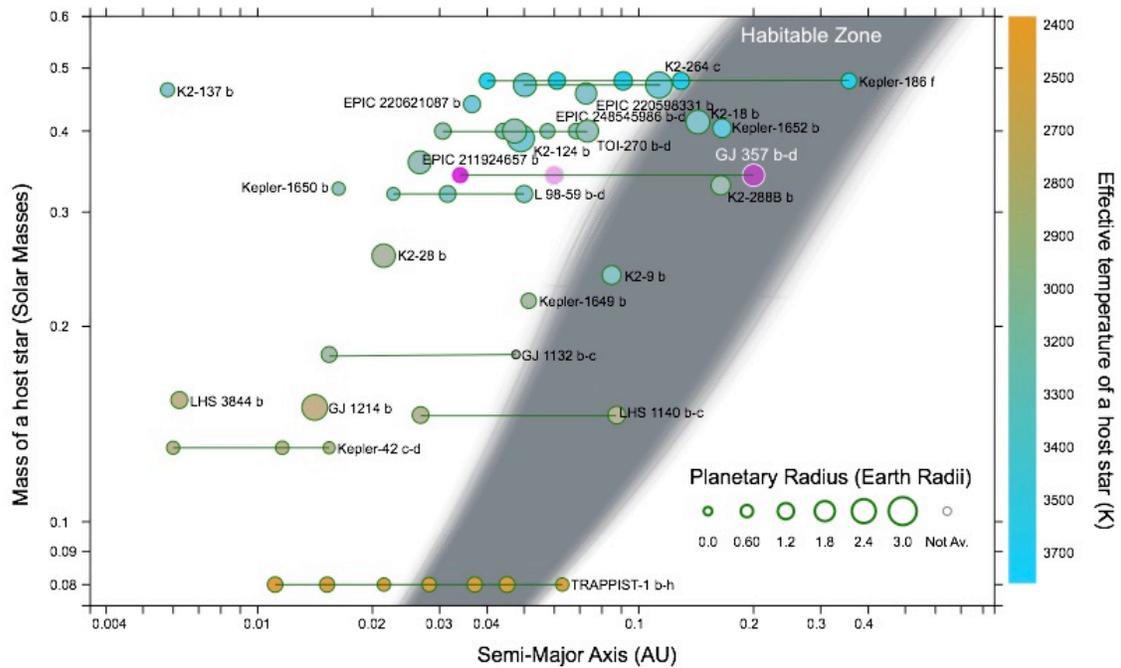

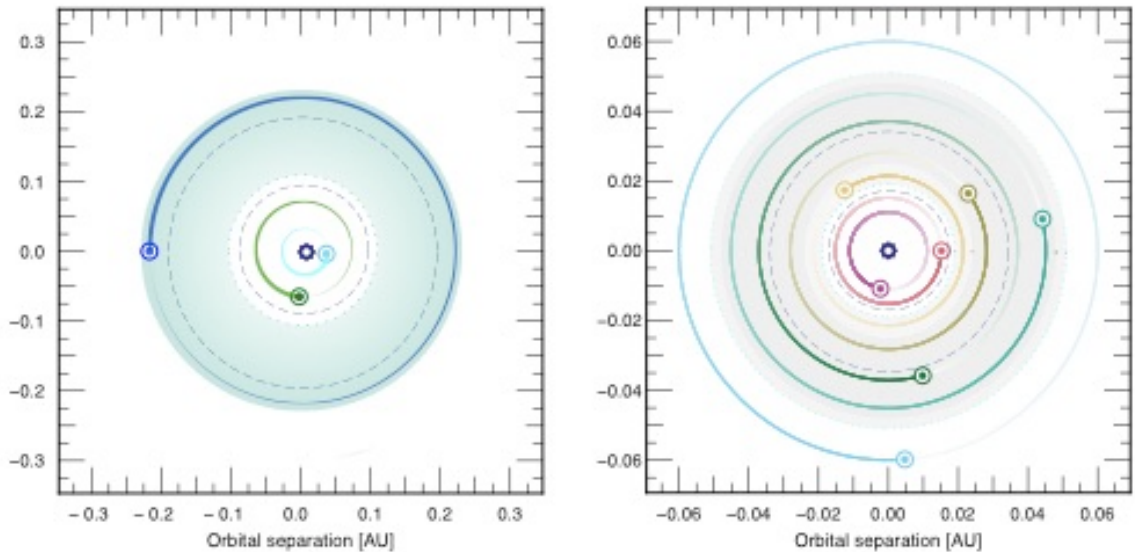

**Figure 2**: (top) Detected M-star planet systems with transiting planets (data exoplanet.eu June 20 2019, radii below 3 $R_{Earth}$), compared to the GJ 357 system – note that only GJ 357b has been detected in transit so far. (bottom) Comparison of the GJ 357 (left, 3 planets) and the Trappist-1 (right, 7 planet) M-star planetary system. The grey shaded region shows the Habitable Zone. Dashed line in both figures on the bottom show the equivalent orbit of present-day Venus and Mars (Trappist-1 figure adapted from Gillon et al. 2017).

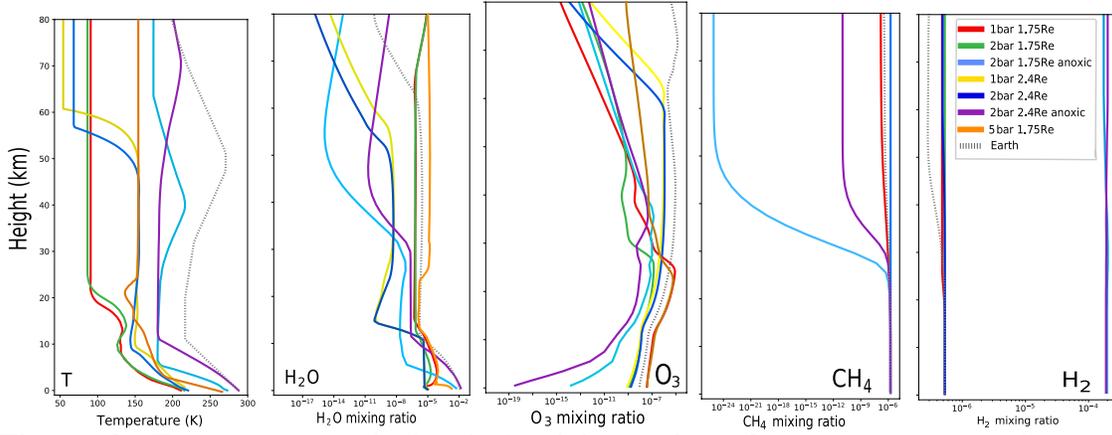

**Figure 3:** Temperature and chemical mixing ratios of water, ozone, methane and hydrogen in the GJ 357 d sample atmosphere models versus height.

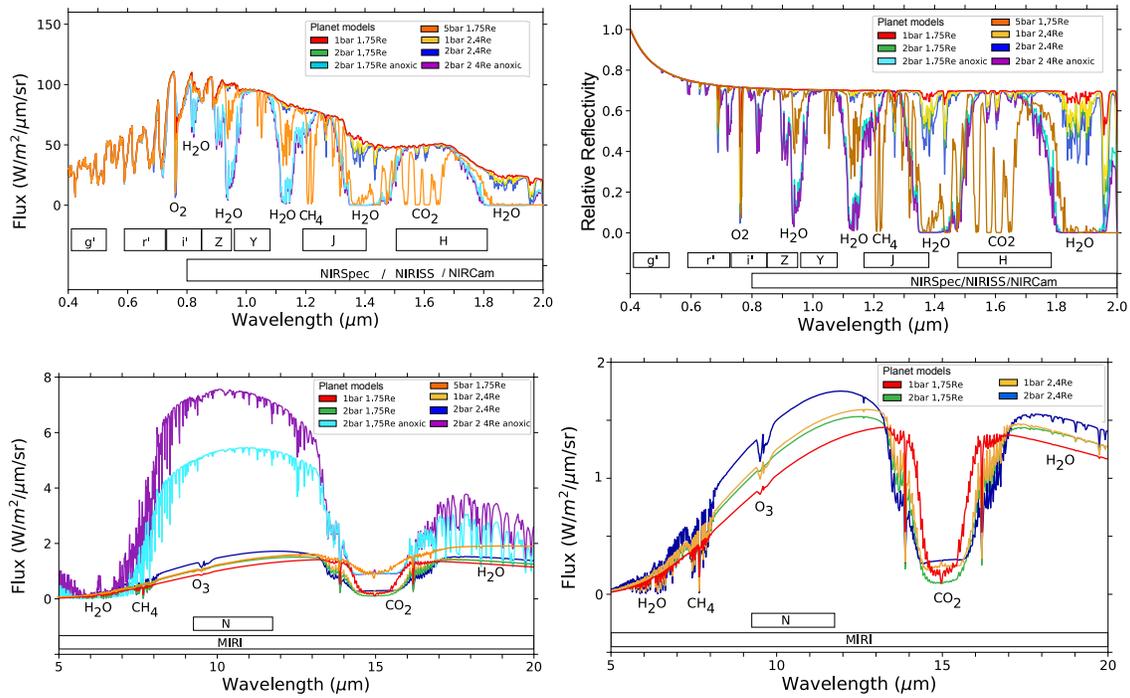

**Figure 4**: Flux of sample model spectra of reflected (top), relative reflection (bottom) and emitted (bottom) of the planet models. Note that the size of the planet is not taking into account and thus the 2.4 and 1.75 $R_{Earth}$ planets show similar flux levels. Spectra data for this figure is available online.

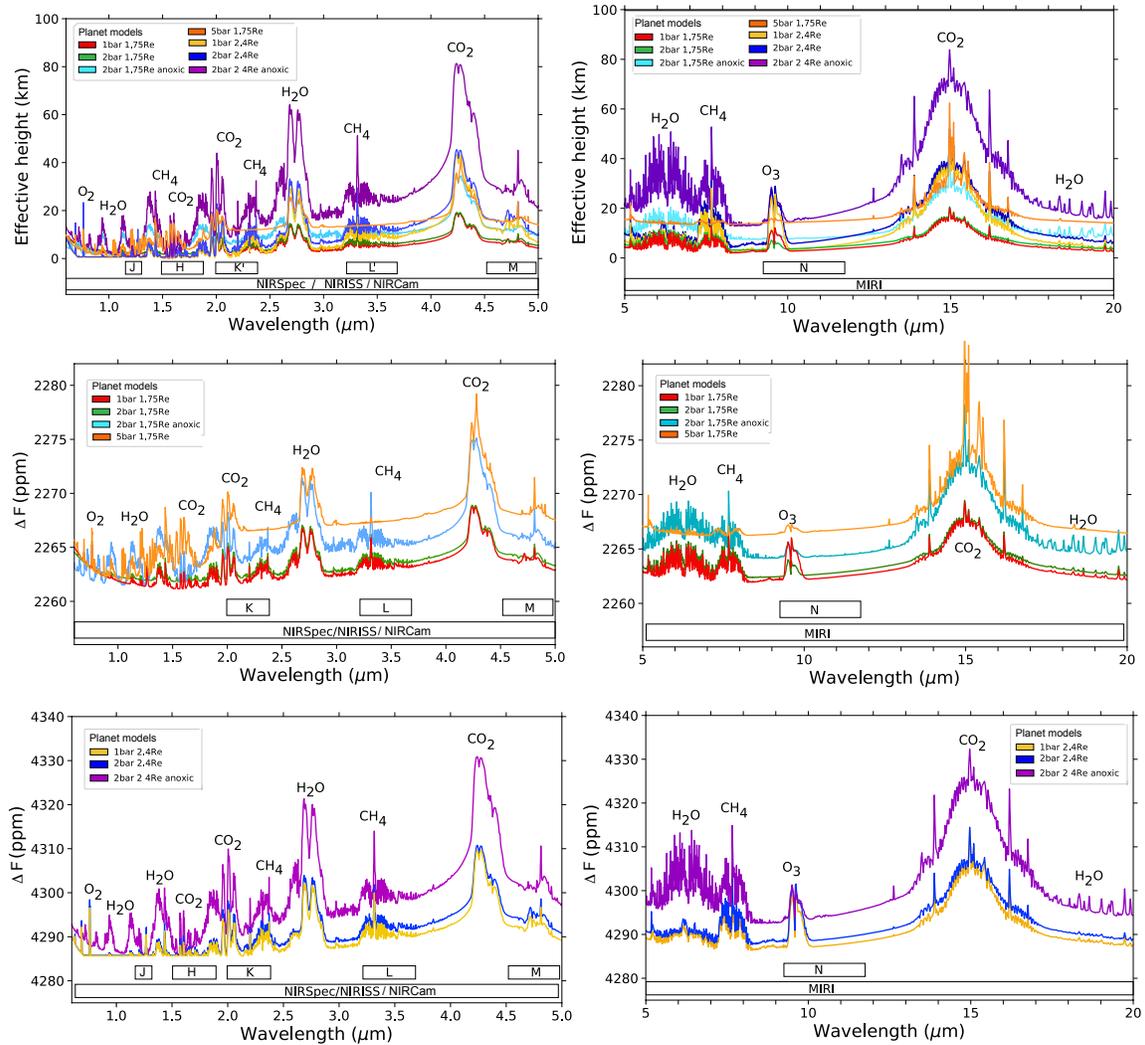

**Figure 5**: Transmission spectra shown in flux difference (Radius planet/radius star)$^2$ in p.p.m. of models of GJ 357 d in the JWST wavelength range. Seven transmission spectra, including four atmosphere models and 2 interior models are shown in (top) effective atmospheric height and flux signal in ppm (middle for 1.75 $R_{Earth}$ models and bottom for 2.4 $R_{Earth}$ models). Spectra data for this figure is available online.

Table 1: Stellar and planetary parameters for GJ 357 system (Luque et al. 2019)

| Star | GJ 357 | | |
|---|---|---|---|
| Name | L 678-39 | GJ 357 | Luyten (1942), Gliese (1957) |
| | TOI 562 TESS Alerts | TIC 413248763 | Stassun et al. (2018) |
| Spectral Type | M2.5V | | Hawley et al. (1996) |
| Brightness | B [mag] 12.52 ± 0.02 | V [mag] 10.92 ± 0.03 | UCAC4 |
| | J [mag] 7.337 ± 0.034 | H [mag] 6.740 ± 0.0332 | 2MASS |
| | G [mag] 9.8804 ± 0.0014 | | Gaia DR2 |
| distance [mas] | 105.88 ± 0.06 | | Gaia DR2 |
| distance [pc] | 9.444 ± 0.005 | | Gaia DR2 |
| Mass [$M_{Sun}$] | 0.342 ± 0.011 | | Schweitzer et al. (2019) |
| Radius [$R_{Sun}$] | 0.337 ± 0.015 | | Schweitzer et al. (2019) |
| L [$10^{-4} L_{Sun}$] | 159.1 ± 3.6 | | Schweitzer et al. (2019) |
| $T_{eff}$ [K] | 3505 ± 51 | | Schweitzer et al. (2019) |
| log g | 4.94 ± 0.07 | | Schweitzer et al. (2019) |
| [Fe/H] | 0.12 ± 0.16 | | Schweitzer et al. (2019) |
| Planets | GJ 357 b | GJ 357 c | GJ 357 d |
| $M_{planet}$ ($M_{Earth}$) | 1.84 ± 0.31 | > 3.40 ± 0.46 | > 6.1 ± 1.0 |
| $R_{planet}$ ($R_{Earth}$) | 1.217± 0.084 | - | - |
| r (g/cm$^3$) | 5.6 ± 1.7 | - | - |
| inclination (deg) | 89.12 ± 0.3 | - | - |
| $t_{Transit}$ (h) | 1.53 ± 0.1 | - | - |
| $a_{planet}$ (AU) | 0.035 ± 0.002 | 0.061 ± 0.004 | 0.204 ± 0.015 |
| $S_{Earth}$ | 12.6 ± 1 | 4.45 ± 0.14 | 0.38 ± 0.01 |

Notes. (a) Error bars denote the 68% posterior credibility intervals. (b) The masses for GJ 357 c and GJ 357 d are a lower limit (Mp sin i) since they are detected from radial velocities only, (d) References. Gaia DR2: Gaia Collaboration et al. (2018); UCAC4: Zacharias et al. (2013); 2MASS: Skrutskie et al. (2006);